\documentclass{jaa}
\usepackage{natbib}
\usepackage{hyperref}
\usepackage[x11names]{xcolor}
\colorlet{purple}{blue!40!red}
\colorlet{bluegreen}{blue!40!green}

\newcommand{\msun}{\ensuremath{\mathrm{M_\odot}}}

\begin{document}\sloppy

%%paper title
%%For line breaks \\ can be used within title
\title{Astrophysics with Compact Objects: An Indian Perspective, Present Status and Future Vision}

\author{ 
Manjari Bagchi\textsuperscript{1,2$^{\dagger}$}, 
Prasanta Bera\textsuperscript{3},
Aru Beri\textsuperscript{4},
Dipankar Bhattacharya\textsuperscript{5,6},
Bhaswati Bhattacharyya\textsuperscript{7},
Sudip Bhattacharyya\textsuperscript{8},
Manoneeta Chakraborty\textsuperscript{9},
Debarati Chatterjee\textsuperscript{6},
Sourav Chatterjee\textsuperscript{8},
Indranil Chattopadhyay\textsuperscript{10},
Santabrata Das\textsuperscript{11},
Sushan Konar\textsuperscript{7},
Pratik Majumdar\textsuperscript{12,2},
%Kuntal Mishra\textsuperscript{10},
Ranjeev Misra\textsuperscript{6},
Arunava Mukherjee\textsuperscript{*12,2},
Banibrata Mukhopadhyay\textsuperscript{13},
Mayukh Pahari\textsuperscript{14},
%Prateek Sharma\textsuperscript{13},
Krishna Kumar Singh\textsuperscript{15,2},
Mayuresh Surnis\textsuperscript{16},
Firoza Sutaria\textsuperscript{17}, \& 
Shriharsh Tendulkar\textsuperscript{7,8},
}
%\affilOne{\textsuperscript{1}Department of P, University X, Place Pincode, Country.}
%\affilTwo{\textsuperscript{2}Department of Q, University Z, Place Pincode, Country.}
\affilOne{\textsuperscript{1}The Institute of Mathematical Sciences, C. I. T. campus, Taramani, Chennai, 600113, India; \footnote{$^{\dagger}$The auther-list is in alphabetical order.} 
}
\affilTwo{\textsuperscript{2}Homi Bhabha National Institute, Anushakti Nagar, Mumbai 400094, India; }
\affilThree{\textsuperscript{3}Department of Physics, Institute of Science, Banaras Hindu University, Varanasi-221005, India; }
\affilFour{\textsuperscript{4} Department of Physics, Indian Institute of Science Education and Research (IISER) Mohali, Punjab 140306, India; }
\affilFive{\textsuperscript{5}Department of Physics, Ashoka University, Sonipat, Haryana 131029, India; }
\affilSix{\textsuperscript{6} Inter-University Centre for Astronomy and Astrophysics, Pune, Maharashtra-411007, India; }
\affilSeven{\textsuperscript{7}National Centre for Radio Astrophysics, Tata Institute of Fundamental Research, S.P. Pune University Campus, Pune 411007, India; }
\affilEight{\textsuperscript{8}Department of Astronomy and Astrophysics, Tata Institute of Fundamental Research, 1 Homi Bhabha Road, Colaba, Mumbai 400005, India; }
\affilNine{\textsuperscript{9}Department of Astronomy, Astrophysics and Space Engineering, Indian Institute of Technology Indore, Indore 453552, India; }
\affilTen{\textsuperscript{10}Aryabhatta Research Institute of Observational Sciences (ARIES), Manora Peak, Nainital, Uttarakhand 263002, India; }
\affilEleven{\textsuperscript{11}Indian Institute of Technology Guwahati, Guwahati, 781039, Assam, India; }
\affilTwelve{\textsuperscript{12,2}Saha Institute of Nuclear Physics, 1/AF Bidhannagar, Kolkata-700064, India; }
\affilThirteen{\textsuperscript{13}Department of Physics, Indian Institute of Science, Bangalore 560012, India; }
\affilFourteen{\textsuperscript{14}Department of Physics, Indian Institute of Technology, Hyderabad, Kandi, Sangareddy 502285, India; }
\affilFifteen{\textsuperscript{15}Astrophysical Sciences Division, Bhabha Atomic Research Centre, Mumbai 400085 India; }
\affilSixteen{\textsuperscript{16}Department of Physics, IISER Bhopal, Bhauri Bypass Road, Bhopal, 462066, India; }
\affilSeventeen{\textsuperscript{17}Indian Institute of Astrophysics, Bangalore 560 034, India; }

%%escape two column mode for title, affiliation and abstract
%%by giving \twocolumn command as shown

\twocolumn[{

\maketitle

\corres{arunava.mukherjee@saha.ac.in}
%%include \corres to print the corresponding author Email id

%%include \msinfo for
%%manuscript information such as
%%received, revised and accepted dates
%%
\msinfo{---}{---}

%%abstract
\begin{abstract}
Astrophysical compact objects, viz., white dwarfs, neutron stars, and black holes, are the remnants of stellar deaths at the end of their life cycles. They are ideal testbeds for various fundamental physical processes under extreme conditions that are unique in nature. Observational radio astronomy with uGMRT and OORT facilities has led to several important breakthroughs in studies of different kinds of pulsars and their emission mechanisms. On the other hand, accretion processes around compact objects are at the core of Indian astronomy research. In this context, AstroSat mission revolutionized spectro-temporal observations and measurements of accretion phenomena, quasi-periodic oscillations, and jet behaviour in binary systems hosting compact objects. Moreover, recently launched XPoSat mission is set to provide an impetus to these high-energy phenomena around compact objects by enabling us to conduct polarization measurements in the X-ray band. Further, during the past decade, numerous gravitational wave signals have been observed from coalescing black holes and neutron stars in binary systems. Recent simultaneous observation of the GW170817 event in both gravitational waves and electromagnetic channels has ushered in the era of multi-messenger astronomy. In the future, synergistic efforts among several world-class observational facilities, e.g., LIGO-India, SKA, TMT, etc., within the Indian astrophysics community will provide a significant boost to achieve several key science goals that have been delineated here. In general, this article plans to highlight scientific projects being pursued across Indian institutions in this field, the scientific challenges that this community would be focusing on, and the opportunities in the coming decade. Finally, we have also mentioned the required resources, both in the form of infrastructural and human resources.
\end{abstract}

%%insert keywords separated by 3 hyphens using \keywords{words}
\keywords{white dwarfs---neutron stars---black holes---multi-messenger astronomy---accretion \& ejection processes---computation \& simulations---astronomical telescopes \& detectors---community \& facility building}

}]
%%close the twocolumn escape here

%%include \doinum{number}for the DOI number in the header
%%include \volnum{number} for the volume number in the header
%%include \year{yyyy} for  year of publication in the header
%%include \pgrange{num--num} page range of article in the header
%%include \artcitid{num} for the article citation id
%%include \lp to print last page of the article
%%include \setcounter{page}{pagenum} for the exact starting page of the article

\doinum{12.3456/s78910-011-012-3}
\artcitid{\#\#\#\#}
\volnum{000}
\year{2024}
\pgrange{1--}
\setcounter{page}{1}
\lp{1}

\section{Introduction}
The study of compact objects spans a wide spectrum of research areas in physics and astronomy: specifically, the general relativity (GR) and strong field gravity, nuclear and ultra-dense matter physics, relativistic magnetohydrodynamics (MHD), thermodynamics and fundamental physics covering all the four fundamental interactions in nature, among many others. 

The compactness, which is a measure of the total mass-energy enclosed in a volume, increases from white dwarfs to neutron stars to black holes. These compact astronomical objects carry observational imprints of relativistic effects of gravity~\citep{Psaltis_2008LRR} and other fundamental forces that appear under extreme physical conditions inaccessible to our terrestrial experiments\footnote{see~\cite{lvk_whitepaper_2024}}. Astronomical observations incorporate detecting electromagnetic radiation from matter around these objects, or the neutrinos emitted by the high-energy matter around these objects, to the extent of detecting gravitational waves that may be emitted due to the merging of these compact objects. Each of these aspects invokes cutting-edge theoretical, experimental and observational methods. In this chapter, we discuss the science cases of white dwarfs (WDs), neutron stars (NSs), stellar mass black holes (BHs), and the astrophysical processes around them in the context of the Indian astronomical community and the open questions that we should focus on in the coming decade.

The organization of this chapter is as follows: in Sections~\ref{sec:accretion}--\ref{sec:blackholes}, we discuss the research directions that are currently being worked on in India and the pertinent open questions in each of the respective fields. In Section~\ref{sec:futureplans}, we discuss future research directions that are ripe for exploration, the resources that would be required to pursue them, and the effort required to build human capital and grow the field.

\section{Accretion \& Jets}
\label{sec:accretion}
Most compact objects (except white dwarfs), being small in their physical extent and less luminous by themselves, cannot generally be detected with their own emission. However, in-falling gas can get heated up significantly by viscous dissipation during accretion onto these compact objects, which could radiate and be detected. An accretion flow, particularly in the strong-gravity around a compact object, is necessarily a nonlinear radiative general relativistic magnetohydrodynamic (MHD) turbulent flow. The accretion flow generally settles into a disk due to its angular momentum and some fraction of the mass and energy is launched from it as a relativistic or non-relativistic jet perpendicular to the disk-plane. Depending on the mass inflow rates, temperatures (which is also related to mass inflow rate) and magnetic fields, the accretion flow undergoes transitions between different dynamic states, classified based on their X-ray spectra as `hard', `soft', `intermediate' etc. \citep[see, e.g.][ for a review]{yuan_narayan_2014_rev}. 

\subsection{Accretion Disks}
Modern observations argue about the imprint of strong magnetic fields in the accretion disk and jet in a black hole system, often termed as a magnetically arrested disk (MAD) or magnetically arrested advective accretion flow (MA-AAF). Often such flows exhibit non-thermal radiation with hard X-rays and gamma-rays. There has been substantial work in the theoretical understanding of the accretion and jet processes, particularly in hard spectral states, through combined modelling of the disk and jet in (non-equilibrium) radiative MHD \citep{mondal_mukhopadhyay_2018, mondal_mukhopadhyay_2019, datta_etal_2022}. Modelling of the interplay between radiation, matter, magnetic and gravitational fields is also related closely to the emergence of quasi-periodic oscillations (QPOs) that are discussed in Section~\ref{sec:qpos}.

\subsection{Formation \& Launching of Jets}
Astrophysical jets from accretion phenomena have been observed, resolved, and studied in great details from radio to X-ray wavelengths~\citep[see e.g.,][]{Begelman1984, Mirabel1994, Fender2004, Ghisellini2001}. However, the exact mechanism by which jets are launched from near the compact object is not well understood. Seminal work by~\citet{blandford_znajek_1977} and~\citet{blandford_payne_1982} suggested that the magnetic field's interaction with the accretion flow along with the effect of black hole's spin (for the former, particularly) can pull and accelerate material perpendicular to the disk plane. Nevertheless, the details of how this actually occurs around compact objects are unclear (see however, e.g.,~\citealt{narayanMAD}). 

\subsection{Disk-Jet Connection}
The quiescent accretion state is a unique stage of accretion when fundamental differences in accreting objects, like the presence or absence of hard surfaces in a neutron star or black hole X-ray binaries, can be probed by spectroscopy~\citep[see ][]{Plotkin2013, Campana2000, Plotkin2013, Parikh2021}. The persistent radio emission during the quiescent state has been detected on several occasions~\citep[e.g.,][]{Gallo2006, dePolo2022}, but its origin is not understood. How a compact object behaves during the ultra-low accretion rate is still far from clear. Few attempts have been made with Chandra spectroscopic studies due to its ultra-low background level and showed interesting features, e.g., soft excess \citep{reis2009} similar to that observed in AGN spectra \citep{done2012,crummy2006,edelson2002,laor1997}. However, the study with Chandra is mainly restricted to 9\,keV, while we expect the emission to dominate in the hard X-ray band ($>$ 7\,keV). No other existing instrument can study the ultra-low accretion state. Therefore, the regime is almost unexplored. 

Previously, jets had only been observed in neutron star and black hole binaries, but new simulations and modeling suggest that jets should exist in cataclysmic variables (CVs) powered by white dwarfs~\citep{coppejans_knigge_2020}. Future observational follow-ups of jets in CVs will be useful for our general understanding of this astrophysical process.

%\subsection{(Spectral) States of Accreting Binaries}
% (b) Why the change of spectral state happen 
 
\subsection{QPOs and Variability}
\label{sec:qpos}
QPOs at various timescales (milliseconds to seconds) seem to be ubiquitous in accreting compact objects, e.g., Type A, B and C QPOs in 0.1-10.0 Hz frequency range are common in BHXBs \citep{casella2005, belloni2002, homan2001} while milliseconds QPOs are common in acceting neutron stars \citep{vanderklis_2000, tod1996}. However, their physical interpretation is contentious \citep{vanderklis_2000, remillard_mcclintock_2006}. QPOs also show up in magnetars and isolated neutron stars~\citep[see e.g.][]{kaspi_beloborodov_2017}, but these are discussed separately below. Also see~\citealt{BMQPO23} for an attempt to explain QPOs in black hole and neutron star sources in a unified manner, and determining spin of black holes based on them. 

There is little understanding as to why and how QPO properties (central frequency, width) are related to the radiative properties (X-ray spectra) and the dynamical properties (e.g. accretion rate, presence/absence of jets, etc). Our understanding of the radiation geometry of accreting microquasars within $\sim 10$ gravitational radii vastly depends on spectroscopic model. The nature of emitting regions closest to the black hole: the innermost part of the accretion disk emitting in X-rays. The jet-base observed in optical/IR/X-ray cannot be easily understood by spectral modelling alone due to its complex nature. The multi-wavelength, correlated, fast variability (of the order of milliseconds) studies have been proven to be an excellent alternative to spectral modelling~\citep{gandhi_etal_2017}. Only a handful of approaches are already successfully providing the quantitative structure of the innermost part of the accretion. The fact that fast variabilities can only originate close to the black hole makes it a robust tool to probe the innermost accreting geometry. However, due to the lack of simultaneous availability of multi-wavelength detectors, only a few efforts were possible in this direction~\citep{ulg2024,vin2019}.

%\section{Population Synthesis \& Studies (1 page)}
%\label{sec:populationsynth}

\section{White Dwarfs}
\label{sec:whitedwarfs}
While binaries containing white dwarfs represent the simplest case of the evolution of compact binaries, there are unresolved discrepancies between the current population models and the properties of the observed samples \citep[e.g.][]{zototovic_etal_2011a, pala_etal_2017}. Current theoretical evolution models are still unable to account for the observed characteristics of the known populations of white dwarfs in both interacting and detached binaries. The observed samples are highly biased, and the population models developed to date have typically been designed to explain the characteristics of sub-samples of these systems, occupying small portions of the vast parameter space. The main objective for the coming decade is to put together a sizable and homogeneous sample of white dwarf binaries that covers the entire spectrum of evolutionary states, to measure precisely their physical characteristics, and to develop further the theory to reproduce the characteristics of the entire population satisfactorily.  

Among the most important aspects of compact binary evolution are the multiple pathways leading to a Type Ia Supernova (SN Ia). The discovery of a range of related thermonuclear explosions, such as SN Iax \citep{foley_etal_2013} and the calcium-rich transients \citep[e.g.][]{perets_etal_2010, perets_etal_2011}, as well as over- \citep{Howell,Filippenko1992a} and under-luminous \citep{Filippenko1992b} SN Ia challenges our understanding of the evolution of close binaries containing white dwarfs. While ongoing and upcoming all-sky high- and low-resolution optical spectroscopic surveys enable us to increase the sample of these systems, high-resolution ultraviolet spectroscopy is crucial for characterising the white dwarfs in these binaries. Planning the next ultraviolet mission must be done immediately because the Hubble Space Telescope is currently the only facility that offers ultraviolet spectroscopy. 

Detailed studies of the physical and atmospheric properties of (non-)interacting white dwarfs in close binaries are only possible in the ultraviolet wavelength, since the contamination from the companion and/or accretion disc is minimal at these short wavelengths \citep{datta2023}. High-resolution ($R \approx 20000 - 40000$) ultraviolet spectroscopy will accurately characterize the white dwarfs in both interacting and detached binaries. Using the extremely precise Gaia parallaxes, effective temperatures and masses of the white dwarfs can be measured by modelling the broad photospheric Lyman absorption lines. Thus, all other quantities that depend on these parameters can be obtained, e.g. mass accretion rates and angular momentum loss rates \citep{Townsley2009}. 

In the case of magnetic CVs, such as AM Herculis, the high X-ray spectral resolution instruments, such as {\it ATHENA}, will open new prospects for plasma diagnostics. Analysing emission line shifts, we will gain a three-dimensional picture of the accretion column, locate the line emission regions, derive parameters of the accretion shock, and directly determine the white-dwarf masses from the gravitational redshift. Repeated observations allow for detailed variability studies of magnetosphere and plasma properties.
 
Fast-timing features, like the presence of QPOs, have been found from a high-density, compact region ($<$1 AU)~\citep{singh2021} during the super-soft phase (SSS) of the recurrent nova~\citep{Kim2020} with very high time-resolution observations in X-rays (thanks to NICER). However, the origin of these QPOs is not yet clear. It is associated with the spin-period; however, low coherence remains unexplained. The short-term variability as observed in X-ray binaries is also found during the SSS phase in a sample of novae. With the advent of the eROSITA survey, many galactic compact candidates (or compact objects in the Magellanic Clouds) are being discovered and will also be discovered in the future. Thus, the time has ripened for the follow-up of X-ray or multi-wavelength observations as they are necessary for the purpose. Another example that compels the requirement of the dedicated multi-wavelength campaigns: Jets are ubiquitous phenomena in accreting sources. Recent works have shown evidence for powerful jets in white dwarfs~\citep{Coppejans2015}, accreting neutron stars~\citep{Migliari2006}, transitional millisecond pulsar systems~\citep{Deller2015}, X-ray pulsars. Therefore, it is the right time to explore similarities and differences between CVs and X-ray binaries. The so-called dwarf novae (DNe), exhibit dramatic outbursts, with recurrence times ranging from weeks to decades similar to that observed in low-mass X-ray binaries (LMXBs) with neutron stars and black holes as the accretors. To date, SS~Cyg is the only CVs with sufficient observations to test the accretion-jet framework. Currently, sensitivity is a key limiting factor, but future generation telescopes such as Square Kilometre Array (SKA) will make a significant difference. SKA will be roughly contemporaneous with other International facilities like Large Synoptic Survey Telescope (LSST), Thirty Meter Telescope (TMT) etc. in optical/IR bands and next-generation successors of Chandra and XMM-Newton such as ATHENA missions. Therefore, multi-waveband observational efforts with wide fields of view will be the key to the progress of this area of transients astronomy.

\subsection{Magnetized \& Spinning White Dwarfs}
%The effects of magnetic fields on the structure and behaviour of compact objects is very challenging to model but in the recent times have led to the observation and modeling of interesting phenomena. 
It is very challenging to model the effects of magnetic fields on the structure and behaviour of compact objects.
However, in the recent times there have been successful efforts leading to modeling the observation and understanding interesting phenomena. AR Scorpii (AR Sco) has been discovered as a close binary system ($P_\mathrm{orb} = 3.56\,\mathrm{hrs}$) comprising a strongly magnetized, rapidly spinning ($P_\mathrm{spin} = 117\,\mathrm{s}$) WD and an M-type main sequence companion star. The system shows pulsed emission over a broad wavelength ranging from radio to X-rays, which has been observed to be modulated over the spin period of the WD and is related to its magnetic interaction with the companion star. Therefore, AR Sco is also dubbed as a WD pulsar because its non-thermal broadband emission appears to be generated by the spin-down energy of rapidly spinning magnetized WD. 

The physical process involved in the pulsed emission from AR Sco, however, is not the same as for traditional isolated neutron star radio pulsars and continues to be elusive. \citet{singh_etal_2020} used multi-wavelength archival data from radio to X-ray observations to model the time-averaged spectral energy distribution of AR Sco under the framework of the leptonic emission model. Theoretical modeling suggests that such systems should also be a faint but detectable source of gamma-rays above 100 MeV, making them good targets for study with the recently inaugurated Major Atmospheric Cherenkov Experiment~\citep[MACE; ][]{singh_2022}.~\cite{BM_ARSCO} also explored the possibility that AR Sco acquired its high-spin and magnetic field due to repeated episodes of accretion and spin-down. An accreting white dwarf can lead to a larger mass and consequently a smaller radius thus causing an enhanced rotation period and a magnetic field, as AR Sco exhibiting, which was confirmed by the authors.

The presence of a strong magnetic field and rotation in a white dwarf can also alter the limiting mass of white dwarfs which is expected to lead to the over-luminous type Ia supernovae (SNeIa). Such over-luminous SNeIa offer the existence of super-Chandrasekhar white dwarfs of mass as high as 2.8\msun, which cannot be explained by the merger of two white dwarfs. 
The magnetic field appears to be one of the exciting viable potential reasons behind such massive white dwarfs (see, e.g., \citealt{BM_PRL, BM_MNRAS15}). 
On the other hand, gravitational wave detection, e.g. GW190814, argues for a massive neutron star of mass 2.6\msun (\citealt{BM_PRD24}, unless it is a very light astrophysical black hole). In either of the massive compact objects, conventional matter therein, i.e. equation of state (EOS), cannot explain them. While EOS for white dwarfs is more or less fixed, EOS for neutron stars, while uncertain at high density due to uncertainties of strong interactions, becomes softer due to the possible emergence of hyperons. Hence, strong magnetic fields could be one of the most potential reasons to make them massive. 

\section{Neutron Stars}
\label{sec:neutronstars}

\subsection{Structure \& Equation of State}
The structure and composition of the neutron stars have remained elusive to date, even after more than half a century since their discovery as radio pulsars in 1967. It remains one of the biggest unsolved problems in astrophysics. The interior of a neutron star harbours the densest form of matter in the observable universe. Neutron stars that are not newly born are cold, degenerate objects compared to the Fermi temperature of ultra-dense state of hadronic matter governed by the strong nuclear interaction, one of the four fundamental forces in nature~\citep{lattimer_prakash_2007}. Thus, understanding the structure and composition of these extreme objects is one of the most important problems that needs to be addressed and resolved satisfactorily in the next couple of decades. The key idea to this understanding is to observationally constrain the theoretically proposed EOS of hadronic matter in the high density, supra-nuclear, cold, degenerate condition along with possible corrections to it due to effects of finite temperature and presence of strong magnetic fields (see~\citealt{Deb2021}). 

Basic nuclear physics predicts several properties of dense matter that can be experimentally measured, e.g., binding energy, charge radii, giant monopole and multipole resonances. Terrestrial experiments at nuclear accelerators and colliders can constrain nuclear interactions of many-body physics to varying degrees. The density at the centre of large nuclei can give us useful constraints on the nuclear symmetry energy and its density dependence along with the behaviour of the iso-spin asymmetry parameter. However, a strong constraint in this context can be achieved for matter densities only up to nuclear saturation density ($\rho_0$), or perhaps a bit higher value ($\approx1.5-2\rho_0$), from the first principle calculations of chiral effective field theory ($\chi$-EFT) and the properties of finite nuclei at terrestrial experiments. On the other hand, the core of a massive neutron star can host hadronic matter at much higher densities, depending on its property, it can go as high as $\approx6-8\rho_0$. At this high density, many of the exotic nuclear degrees of freedom become theoretically possible. For example, the existence of a quark star or hybrid star, hadron-to-quark phase transition, the existence of hyperonic matter, strange quarks, and various types of condensates of mesons are some of the most intriguing questions that need to be answered. This is a completely new, unexplored domain of extreme matter physics that is inaccessible experimentally other than astronomical observations of certain properties of neutron stars. Another related point is superfluidity and superconductivity. At such high densities, proton Cooper pairs are expected to form which offers the possible superconducting phase of a neutron star, depending on its internal temperature and underlying critical temperature. 

Observations of various stellar structural properties of neutron stars, e.g., mass, radius, tidal deformability, moment-of-inertia, quasi-normal mode, can provide us with a promising direction to understand the above mentioned states of matter~\citep{lattimer_prakash_2007}. The most widely adopted view in this regard is to exercise observational constraints on theoretically proposed EOS of the neutron stars. The global structural properties of the neutron star, such as mass, radius, tidal deformability and moment-of-inertia of neutron stars are governed by the equations of state as predicted from their stable solutions of Tolman-Oppenheimer-Volkoff (TOV) equation for hydrostatic equilibrium in general relativity. Recent observations of binary neutron star merger GW170817 event have strongly constrained a number of stiff EOS~\citep{gw170817_discovery_2017, gw170817_eos_2018, lattimer_2019} which has been predominantly estimated using relativistic mean field theory framework. Subsequently, we observed further constraints on neutron star EOS from X-ray observations of hot-spots on the surface of a couple of neutron stars~\citep{miller_etal_2019, raaijmakers_etal_2019}. Combined with the multi-messenger observations of these neutron star properties with the experimental constraints from the physics of finite nuclei, it can provide a tighter constraint on the EOS \citep[e.g. ][]{raaijmakers_etal_2020}. Several recent studies have also combined multi-disciplinary information from nuclear theory calculations of Chiral effective field theory at low densities, heavy-ion collisions at intermediate densities and multi-messenger astrophysical data and perturbative QCD calculations at high densities to impose constraints on the dense matter EOS~\cite{Ghosh_2022, GhoshPradhan_2022, ShirkeGhosh_2023}.
X-ray observations have significantly advanced our understanding of neutron star structures and EOS of ultra-dense matter. Instruments like NASA's Neutron Star Interior Composition Explorer (NICER)~\citep{NICER} have provided precise measurements of neutron star masses and radii, offering critical constraints on theoretical EOS models. For instance, NICER's observations of pulsars PSR J0030+0451 and PSR J0740+6620 have yielded mass-radius estimates that challenge existing EOS predictions, suggesting that neutron star matter is less compressible than previously thought~\citep{miller_etal_2019, Riley_etal_2019, Miller_2021, Riley_2021, Choudhury_2024}. \\

Despite these advancements, current X-ray instruments face limitations (such as detecting fainter sources) that hinder a comprehensive understanding of neutron star interiors.~The limited sample size of neutron stars further constrains the generalization of findings across the neutron star population. Many precise mass measurements rely on neutron stars in binary systems, which may not fully represent isolated neutron stars with potentially different structural properties. To address these challenges, instruments such as {\it eXTP}~\citep{Zhang2019}, {\it STROBE-X}~\citep{Paul2019} with larger collecting area and higher time resolution would significantly enhance sensitivity, enabling more precise measurements of neutron star properties and better constraints.

For neutron stars, X-ray polarization provides valuable insights into the structure of the powerful magnetic fields present on their surfaces. The degree and angle of polarization can help determine the configuration of these magnetic fields, revealing whether they are dipolar or have a more complex structure. X-ray polarization measurements, particularly in the soft X-ray band, allow us to explore the thermal emission from the surface of neutron stars. This thermal emission is closely tied to the neutron star’s EOS. By analyzing the polarization of this emission, we can gain a deeper understanding of the internal composition and structure of neutron stars.

\subsection{Pulsars}
A sub-sample of neutron stars emits beams of electromagnetic radiation that make them detectable as pulsars. Their spin period is very stable and so the pulses detected as they spin are like the ticks of an extremely precise clock. Pulses from the pulsars are detected only when their beam of emission is pointed toward Earth. The fastest rotating neutron stars with a spin period of less than 30 ms are called millisecond pulsars (MSP). The presently known population of about 3400 pulsars is less than 5\% of the predicted number of detectable radio pulsars \citep[e.g.][]{faucher06}, indicating that a lot of new pulsars are going to be discovered with existing and upcoming sensitive observing facilities as well as with improved analysis techniques. Recently the blind and targeted surveys using the Giant Meterwave Radio Telescope (GMRT) have resulted in the discovery of pulsars including millisecond pulsars, very wide profile pulsars, pulsars with gamma-ray counterparts, etc (GHRSS survey, Bhattacharyya et al. 2016, 2019) and the Fermi directed targeted survey~\citep{bhattacharyya2013gmrt, Bhattacharyya2021, bhattacharyya_etal_apj2022, Roy_2015}. 

Pulsars can be used as remarkably precise ``celestial clocks'' to explore many different aspects of physics and astrophysics. Precision pulsar timing is the exploration of this concept using the measurement of a sequence of pulse time of arrivals (ToA) ranging from hours to decades. Being extremely stable rotators, MSPs act as laboratories for the study of matter in extreme conditions. Long-term timing of the MSPs can probe these precise timekeepers. As examples, an ensemble of pulsars has been used to detect low-frequency gravitational waves \citep{PTA}. They can be used to put limits on alternative theories of gravity~\citep{altgrav}, dense matter equations of state~\citep{radioEoS1, radioEoS2}. Moreover, as~\citep{skumari2022} published a decade of timing study for a MSP discovered by the GMRT that has allowed the team to get a precise estimation of the velocity of this MSP as well as to study its orbital behaviour and properties of the intervening medium. The equations of state of cold nuclear matter can be probed using reliable estimates of neutron star mass. However, such measurements possible with the timing of millisecond pulsars at radio frequencies are rare. The recent effort of neutron star mass estimates from gamma-ray eclipses in compact millisecond pulsar binaries (which were discovered in Fermi-directed radio searches) using the Fermi-gamma ray telescope~\citep{clark2023} provides useful input for that. This is an example where the synergy of radio observations (where GMRT also participated) along with the Fermi-gamma-ray space telescope allowed the team to reach the science goal.  

Recently, a number of pulsars have been discovered below the conventional death line~\citep{deathline}, challenging our current understanding of pulsar radio emission mechanism. We might expect more such `below death line' pulsars to be discovered in the future with SKA. Study of their properties observationally and theoretically will improve our understanding of radio pulsar emission mechanisms and their turning off.

\subsection{Accretion and Nuclear-Powered Millisecond Pulsars in Low Mass X-ray Binaries}

Rapidly rotating neutron stars (NS) are also found in binary systems with a low-mass companion star (less than one solar mass) and have relatively weak magnetic fields, typically in the range of $\sim 10^8-10^9$ Gauss. Many of these systems are in globular clusters, old star clusters that allow enough time for magnetic fields to decay from the much stronger fields seen in younger neutron star (usually above $10^{12}$ Gauss) to significantly weaker levels. As a result, it was suggested that old neutron star could be spun up to millisecond periods through the accretion of matter and angular momentum during a Low Mass X-ray Binary (LMXB) phase. This process is known as the recycling scenario~\citep[see e.g.,][]{Bhattacharya1991}.~These weakly magnetized, accreting NSs in LMXBs can be spun up to rates of several hundreds of Hertz~\citep{Alpar82} and are known as accreting millisecond X-ray pulsars~(AMXPs). So far, only 25 AMXPs \citep[see, e.g.,][]{patruno2012,Bult2022,Ng2022,Beri2023a} have been found, and the discovery of these systems forms an important missing link in the recycling scenario~\citep[see e.g.,][]{Salvo2020, Alessandro2022}. Courtesy to the availability of detectors in radio, optical, X-ray, and Gamma-ray bands, the recycling scenario has also been investigated in the multi-wavelength observations: the evolutionary path leading to the formation of a Millisecond-spinning Pulsar. \\

In addition, there exist nuclear-powered X-ray millisecond pulsars (NMXPs)~\citep[see e.g.,][]{Strohmayer1998, Strohmayer1999, Strohmayer2001} in NS-LMXB systems. In NMXPs coherent millisecond period brightness oscillations have been observed during thermonuclear X-ray bursts (sudden eruptions in X-rays, intermittently observed from NS-LMXBs). Besides, there exists a partial overlap between AMXPs and NMXPs which implies, some AMXPs are also NMXPs and vice versa~\citep[for example, SAX J1808.4--3658, IGR J17511--3057, HETE J1900.1--2455, XTE J1739--385][]{Chakrabarty2003, Strohmayer2003, Altamirano2010, Sudip2010, Bilous_2019, Beri2023a}. Currently, only 19 confirmed NMXPs~\citep[see][for a recent review]{Sudip2021} are known. \\

Therefore, to be able to hunt for more such systems, we need timing and spectroscopic instruments with large effective area not only in X-rays, but also in UV and optical bands to be able to detect any short time-scale stochastic and/or periodic UV-variability and correlate this against the variations seen at optical/X-ray wavelengths. 

\subsection{X-ray Pulsars}

One of the key observational signatures in studying the magnetic fields of these neutron stars is the detection of cyclotron resonance scattering features (CRSFs) in their X-ray spectra \citep[see e.g.,][]{Maitra2018,Varun2019, Beri2021,Ashwin2024}. These features are directly linked to the strength of the magnetic field and result from the modification of the Compton scattering cross-section under the influence of intense magnetic fields. The X-ray polarization observed in these pulsars, particularly when resolved by the pulse phase, is highly sensitive to the topology of the neutron star’s magnetic field due to this effect.

Theoretical models predicted a high polarization degree (PD) of approximately 60-80{\%} in these systems. However, recent observations by {\it IXPE} ({\it Imaging X-ray Polarimetry Explorer}) have revealed a surprisingly lower PD of 10-20{\%}, first observed in pulsar Her X-1~\citep{Doroshenko2022} and subsequently in other pulsars. Although the cause of this discrepancy remains uncertain, it is speculated to be related to the properties of the neutron star's atmosphere.

{\tt POLIX} on-board {\it XPoSat}~(India's first dedicated X-ray polarization instrument\footnote{World's second X-ray polarization mission}), with its sensitivity to higher energy photons compared to {\it IXPE}, is expected to provide crucial insights into this issue. Given that the pulsed fraction—representing the relative strength of pulsations—increases with energy, {\tt POLIX} could potentially detect the anticipated polarization enhancement near the CRSF. This feature, typically found around 30-40 keV, has not been observed before in polarization studies.

Laboratory experiments and simulations suggest that {\tt POLIX} could indeed observe this polarization enhancement in bright sources, which would offer significant contributions to our understanding of magnetic field configurations in these systems.~After the launch of {\it XPoSat}, several X-ray pulsars have been observed including Crab and the dedicated teams are analyzing these data and heavily involved in understanding any background contributions. Furthermore, the combination of broadband X-ray polarization and X-ray interferometry could greatly enhance the study of CRSFs by resolving the specific regions on the neutron star’s surface where these features originate, providing a detailed map of the magnetic field structure.

\subsection{Ultra-luminous X-ray Sources}
Ultra-luminous X-ray sources (ULXs) are point-like, off-nuclear objects with X-ray luminosities exceeding the Eddington limit for a $\sim 10$-solar-mass black hole ($Lx > 10^{39} ergs/sec$). Initially, these sources were believed to be intermediate-mass black holes (IMBHs) accreting at sub-Eddington rates. However, the recent discovery of neutron stars within some of these extra-galactic ULXs has confirmed that super-Eddington accretion can occur in certain cases~\citep[e.g.,][]{Bachetti2014, Fürst_2016}. This finding has sparked an ongoing debate about the true nature of the compact objects driving ULXs and producing such extreme X-ray luminosity. \\

ULXs are often located in crowded fields, making it difficult to distinguish them from nearby sources. Therefore, advanced imaging capabilities with high angular (sub-arcsecond) resolution, similar to that of the Chandra X-ray Observatory, are essential for resolving ULXs from confusing nearby sources. The discovery of pulsations in the X-ray data of ULX M82 X-2 revealed that ULXs can be powered by NS. This has led to the identification of a new subclass known as ULX pulsars. Given that only a few ULX pulsars have been identified so far, conducting detailed studies of known systems and searching for new ones is crucial to advance our understanding of these objects. An instrument with fast timing and monitoring capabilities would be critical in the search for new ULX pulsars. \\

Additionally, detecting proton cyclotron line allows magnetic field measurement in these systems. 
This will help verifying the idea if they are highly magnetized stellar mass black hole sources accreting at the sub-Eddington rate~\citep{mondal_mukhopadhyay_2019}. ULXs exhibit several spectral features that are important for understanding the accretion geometry in these systems. These include a spectral cut-off around 6-10 keV, an emission line near 0.9 keV (suggesting the presence of outflows), and a hard X-ray tail \citep[see e.g.,][]{Jaisawal2019, Beri2020,Lin2022, Birendra2024}, all of which require further investigation to determine their origin. In order to accurately constrain these features, observational facilities equipped with broadband X-ray coverage (from sub-keV to $\sim 100$ keV) of high sensitivity, good energy resolution, and millisecond timing resolution is essential.

%{X-ray pulsars, type-I bursts and BO should also be mentioned. X-ray polarization would be important to discuss.} 

%Although inclinations can be inferred from subtle features in optical light curves, such estimates may be systematically biased due to incomplete heating models and poorly understood variability.   
%systems MSPs often have orbital companions. In some of the MSP systems, (called “spider MSPs”) the pulsar and the
%companion star could have separations as small as that between the earth and the moon.

%\subsection{Equation of State}

\subsection{Physics and Astrophysics with Gravitational Waves}

Neutron stars are one of the most suitable tools to study gravitational physics. When neutron stars are binary radio pulsars, they can be used to test various aspects of general relativity and put better constraints on alternative theories of gravity. Stable millisecond pulsars can be used to detect nano-Hz gravitational waves and this effort is already ongoing by the Indian Pulsar Timing Array consortium as a member of the International Pulsar Timing Array collaboration. Hence, more synergy between neutron star researchers with the researchers working on gravitational physics is encouraged.

One of the primary goals in this research area is to focus on various aspects of nuclear physics in the context of astronomy and astrophysics. In particular, there has been an effort to pursue the measurements of nuclear matter properties of the dense, degenerate state of hadronic matter at the core of the neutron stars~\citep{Forbes_etal_prd2029, Dietrich_etal_2020sci, Capano_etal_NatAs2020, Huth_etal_2022Nat}. Recently gravitational wave observation of the GW170817 event has enabled us to put strong constraints on the neutron star EOS~\citep{gw170817_discovery_2017, gw170817_eos_2018, miller_etal_2019, Riley_etal_2019, raaijmakers_etal_2019}. In the future, as the sensitivity of GW detectors increases, we will observe many such events of binary neutron star (BNS) mergers and will detect them with larger signal-to-noise ratios (SNRs)~\citep{ET_science_2020JCAP, CE_science_2023arXiv230613745E}. Several groups in India are involved in incorporating state-of-the-art constraints from multidisciplinary physics at different density regimes (nuclear theory, nuclear and heavy-ion experiments, multi-messenger astrophysics) within a Bayesian approach to get a unified picture of nuclear interactions~\citep{Ghosh_2022, GhoshPradhan_2022, ShirkeGhosh_2023, Biswas_2021, Biswas_2022, Imam_2022, Imam_2024, Malik_2022, Patra_2022, Patra_2023, Venneti_2024, roy2024}.

Continuous gravitational waves (CW) are another class of GW signals that the LIGO-Virgo-KAGRA (LVK) community is actively searching for but hasn't detected yet~\citep{KeithRiles_LRR_2023, Wette_review20ycw_2023APh}. Detecting these signals from accreting neutron stars in low-mass X-ray binary systems (e.g., Sco X-1) is a prime objective~\citep{MMR_spin_wandering, kar_etal_apj2025}. An active effort is in place to detect these CW signals from X-ray bright sources, e.g., Sco X-1~\citep{lvk_scox1_hmm_viterbi, lvk_scox1_2022ApJL, MPW_PRD2023}, as well as known pulsars in our galaxy~\citep{lvk_2025arXiv250101495T}. However, searching large volume of unknown source parameters is a computationally challenging barrier that we need to overcome in the future~\citep{MPW_PRD2023, Wette_review20ycw_2023APh}. Gravitational wave emissions from unstable oscillations in neutron stars also contain signatures of their internal composition and can also be important probes of the EOS. There have been several recent investigations on the effect of the nuclear EOS in neutron star quasi-normal modes such as $f$-modes~\citep{Pradhan_2021, Pradhan_2022, Pradhan_2023, Pradhan_2024, Shirke_2023, mondal_bagchi_2024} or $g$-modes~\citep{TranGhosh_2023}, which may be excited during inspiralling binaries or $r$-modes~\citep{GhoshPathak_2023, GhoshMNRAS_2023} which are possible sources of CW emission.

%\subsection{Prospects of Multimessenger Astrophysics} 
%
%{\color{blue} Neutrinos from supernovae, GRBs, AGNs and connection to other astroparticle physics should at least be briefly mentioned. }

\subsection{Understanding the population of neutron stars}
Neutron stars are observable in various forms, e.g. accreting X-ray binaries, radio pulsars, magnetars, X-ray dim isolated neutron stars (XDINSs), etc. However, it is not known what fraction of each of these populations has been discovered and what fraction is yet to be discovered. Hence, a detailed population synthesis study will be necessary. It will also be good to know which part of the Galaxy harbour more undiscovered neutron stars.

Several aspects of transient astronomy depend on our understanding of compact object binaries. One crucial aspect of this is related to the details of the supernova physics which determines the birth mass function and natal kicks the compact objects receive. Uncertainties also remain in creating a metallicity-dependent map between compact objects and their progenitors. Distinguishing stellar BHs from NSs is another crucial issue, and related to this, it is important to constrain the upper mass limit of NSs and lower mass limit of BHs, and whether there is a mass gap between the two.

\subsection{Evolutionary links among different neutron star classes}

Today, the number of neutron stars discovered is $\gtrsim$3500  with a wide variety of observational characteristics.  Yet, they belong to three basic categories - a) the rotation-powered pulsars (RPP); b) the accretion-powered pulsars (APP)  and c)  the internal energy powered (IEP)  objects where the emission comes from certain internal reservoirs of energy like the post-formation residual heat or energy stored in ultra-strong magnetic fields.  These categories are neither mutually exclusive, nor are their evolutionary connections unknown. One of the prime challenges of neutron star research has always been to find a unifying theme to explain the menagerie of disparate observational classes.

The connection between the RPP and  APP has been studied through the decades and is quite well understood.  The latest focus on this line of investigation is the development of a detailed theory of magneto-thermal evolution and interconnections between different types of isolated neutron stars, where the target objects include radio pulsars and most of the IEP  class objects such as magnetars, the central compact objects (CCOs), and the X-ray dim Isolated Neutron Stars.

Observational evidence for evolutionary connections between the IEPs and radio pulsars has also started accumulating of late.  A number of slow radio pulsars, with periods as large as or larger than typical IEPs  (like,  J1903+0433,  $P_s$  = 14.05s  and  J0250+5854,  $P_s$  = 23.54s),  have been detected only with advanced observational capabilities~\citep{Khargharia_etal_apj2012, Kou_etal_apj2019, Agar_etal_mnras2021}.   Magnetar-like X-ray bursts have been detected from high-magnetic field radio pulsars, and some of the Magnetars have also been observed to emit periodic radio pulses.  Clearly,  the boundary between the RPPs and the IEPs is getting blurred. The need of the hour is to develop robust theoretical models to describe these interconnections and the evolutionary pathways.

The nature and the evolution of the magnetic field, ranging from $10^8$~G in millisecond pulsars to $10^{15}$~G  in magnetars,  is an important factor shaping such evolutionary pathways.  It plays an important role in determining the evolution of the spin, the radiative properties, and the interaction of a neutron star with its surrounding medium. Consequently, understanding the nature of the magnetic field, particularly in high-field objects, is also of great importance in today's neutron star research.

\section{Black Holes}\label{sec:blackholes}
\subsection{Spin: A Window into Black Hole Dynamics}
Black holes remain one of the most enigmatic and captivating phenomena in astrophysics, pushing the boundaries of our understanding of fundamental physics. The angular momentum $J$ of a black hole having mass $M$, can be defined in terms of its dimensionless spin $a=Jc/GM^2$. This provides important input into the theoretical understanding of the endpoints of stellar evolution and the asymmetric supernova kicks that create such misaligned systems. Moreover, black hole spin -- a measure of its angular momentum -- plays a crucial role in shaping the behavior of the surrounding accretion disk, driving energetic outflow-like relativistic jets, which are of great wider relevance since feedback onto the surroundings through expulsion of matter is believed to allow supermassive BHs to regulate galaxy formation and influence star formation by either quenching it through gas expulsion ~\citep[see e.g.,][]{Matteo2005,Springel2005,Dubois2012} or triggering it via shock compression and turbulence ~\citep[see e.g.,][]{Silk1998, Gaibler2012, Beckmann2025}. A new mission concept {\it High Energy X-ray Probe} ({\it HEX-P}; 0.2-80~\rm{keV}) has also been proposed to NASA to probe super-massive BHs growth and driver of Galaxy evolution~\citep[see][for details]{Garcia2024, Connors2024}. \\

Observations from advanced X-ray missions, such as {\it NuSTAR} and {\it XMM-Newton}, have provided significant breakthroughs in measuring spin of black holes in X-ray binaries. These measurements rely on the detailed analysis of reflection features arising from the innermost regions of the accretion disk, where relativistic effects are most pronounced. The iron K$\alpha$-line at 6.4 keV is a particularly powerful diagnostic, as its broadening and skewing \textendash{} caused by gravitational redshift, Doppler and light bending effects near the event horizon \textendash{} offer a direct window into estimation of spin of the black hole. Additionally, Compton reflection hump, a feature peaking at $\sim 20$--$30$ keV, provides complementary information about the inner disk radius and the intensity of X-ray reflection, which are influenced by spin. Several studies have been performed that have significantly advanced theoretical and observational techniques for extracting spin parameters from X-ray spectra, underscoring the critical role of high-resolution spectroscopy in refining our understanding of black hole accretion physics~\citep[e.g.,][]{Reynolds2014, García_2014}. 

Nonetheless, uncertainties arising from assumptions about the accretion disk structure-such as deviations from a thin, Keplerian disk-necessitate improved modeling and observational techniques. The complication arises due to complex hydrodynamic and radiative processes active in these systems. While the Indian community, by and large, have been using available codes to describe such complex spectra and testing them on {\it AstroSat} and other satellite data, there is a potential for the community to significantly contribute to the creation of next generation of those codes. 

The potential development of X-ray interferometry~\citep{Uttley2021} could mark a groundbreaking leap in our ability to study fine details in the high-energy universe. Once realized, this technology would achieve resolutions as fine as micro-arcseconds, potentially enabling direct imaging of regions near a black hole’s event horizon. Similar to the success of {\it Event Horizon Telescope} in the radio spectrum, X-ray interferometry would operate at much higher energies, offering the chance to observe entirely different physical processes. By resolving the iron K$_\alpha$ line-profile from the accretion disk, it could provide detailed measurements of velocity and gravitational redshift as functions of position, paving the way for precise determinations of black hole mass and spin. 

A critical aspect of an X-ray interferometer is its optical system. X-ray mirrors require grazing incidence reflection. Recent advancements in lithographic techniques (also used in {\it {ATHENA}} optics) enable the fabrication of high-precision flat mirrors, including slatted mirrors that facilitate the parallelization of interfering beams~\citep{Willingale2004}. Another key factor in its success is imaging sensitivity, requiring large collecting areas and highly efficient detectors. Current technologies, such as Transition Edge Sensors (TES) (also proposed for Lynx~\citep{Bandler2019}), provide exceptional energy resolution ($\sim 2$ eV), but they require cryogenic cooling, adding complexity to space missions. An alternative approach could involve developing novel one-dimensional (1D) pixel arrays, reducing the need for extensive cryogenic systems while maintaining high-resolution capabilities in the long term; beyond 30 years, India could aim for a full-scale, multi-spacecraft X-ray interferometry mission capable of achieving sub-micro-arcsecond resolution.  

\subsection{Relativistic Jets: its origin and production mechanism of the non-thermal components}

Relativistic jets, streams of highly energetic particles ejected from the vicinity of black holes, are among the most dramatic astrophysical phenomena. These jets are believed to originate from the interplay between the black hole’s spin and its magnetic field, with two competing mechanisms proposed: the Blandford-Znajek process that links jet formation to the extraction of rotational energy from the black hole~\citep{blandford_znajek_1977}, and the Blandford-Payne process that attributes jet production to disk winds~\citep{blandford_payne_1982}. X-ray interferometry offers a promising avenue to directly resolve the jet base and its connection to the accretion disk, potentially distinguishing between these mechanisms. Observations of the velocity structure and magnetic fields in the jet-launching region could illuminate the processes driving relativistic jets, addressing long-standing questions about their origins and energetics.~Simultaneous multi-wavelength observations at regular and high cadence of sources (especially those with jets) need to be undertaken in a coordinated manner. With a number of meter-class optical telescopes, {\it GMRT}, {\it AstroSat} and future observatories, India can make a significant contribution in this area. Our involvement in the {\it Cerenkov Telescope Array} ({\it CTA}) would also be crucial for such an endeavour.

\subsection{X-Ray Variability: nature and geometry of accretion flows near the event horizon}
The variability of X-ray emissions from black hole binaries across a wide range of timescales -- from milliseconds to years --provides a direct window into the dynamic processes occurring near the event horizon. QPOs are particularly significant as they represent oscillatory phenomena within the accretion disk, offering potential diagnostics of black hole spin, disk structure, and the interaction of the disk with the black hole’s relativistic environment. Low-frequency QPOs, typically observed at 0.1–30 Hz, are associated with fluctuations at the inner edge of the accretion disk, while high-frequency QPOs, in the range of $40-450~Hz$, are thought to correspond to Keplerian motion at the innermost stable circular orbit (ISCO). 

Instruments such as the {\tt PCA} onboard {\it RXTE} satellite was foundational in timing studies, leveraging its high temporal resolution and large effective area to detect weak and transient QPO signals. It enabled precise power spectral analysis, revealing the presence of multiple QPO modes and their correlations with spectral states in black hole binaries. Building on these foundations, {\it AstroSat} has significantly contributed to our understanding of X-ray variability with its {\tt LAXPC} and {\tt SXT}. 

{\tt LAXPC}, which is sensitive in $3-80$~keV energy range, provides high temporal resolution and excellent sensitivity for bright sources, making it particularly suited for QPO studies. For example, \citet{Yadav2016} used {\tt LAXPC} to study QPOs in GRS 1915+105, exploring their energy dependence and correlations with X-ray spectral states. Similarly, \citet{Misra2017} examined variability in Cygnus X-1, leveraging {\tt LAXPC} data to analyze spectral-timing correlations. The {\tt SXT}, covering the 0.3–8 keV range, complements these observations by providing broadband spectral coverage and allowing simultaneous studies of variability and spectral features. \citet{Singh2017} demonstrated the utility of {\tt SXT} in mapping low-energy variability in black hole binaries.

Advanced techniques such as time-lag analysis and frequency-resolved spectroscopy have further enhanced the insights gained from {\it AstroSat} data. Time-lag analysis, used to measure delays between correlated signals in different energy bands, has been instrumental in identifying physical origins of variability, such as disk-corona interactions or relativistic light-bending effects. Frequency-resolved spectroscopy, which disentangles the spectral contributions of different variability components, has allowed for detailed studies of how changes in the accretion flow propagate through the system, providing a deeper understanding of the geometry and dynamics of accretion flows \citep[see e.g.,][]{Revnivtsev1999, Gilfanov2003, Uttley2011, Ingram2012}. Despite these advancements, {\it AstroSat}’s sensitivity is limited, particularly for the faint sources or weak QPOs. This limitation underscores the need for next-generation instruments with higher sensitivity and improved imaging capabilities. Moreover, to obtain a comprehensive understanding of the accretion geometry, we need simultaneous multi-wavelength observations. For example, recent studies on the black hole candidate Swift~J1357$-$0933 have provided critical insights into variability mechanisms in transient systems~\citep{Beri2019, Paice2019, Beri2023b}. 

\subsection{X-Ray Transients: Tracking State Transitions}

X-ray transients, which exhibit dramatic changes in luminosity and spectral properties, provide a unique opportunity to study the transitions between different accretion states in black hole binaries. These transitions are often accompanied by shifts in the geometry of the accretion disk and the corona, offering crucial insights into the physical processes governing these systems~\citep{2011MNRAS.418..490C}. Disk instability models, such as the ``inside-out'' or ``outside-in'' scenarios, have been proposed to explain the onset of outbursts~\citep{Dubus2001}, but observational limitations often hinder a detailed understanding of these phenomena. 

Multi-wavelength campaigns combining X-ray, optical and ultra-violet observations have been proven essential for capturing the evolution of transients. The XB-NEWS monitoring program, which uses optical observations to identify potential transient sources, has been instrumental in enabling follow-up observations with {\it AstroSat}. With its suite of instruments such as {\tt LAXPC}, {\tt SXT}, and {\tt UVIT}, {\it AstroSat} provides a powerful platform for studying transients across multiple wavelengths. The combination of optical monitoring from XB-NEWS and rapid-response X-ray follow-up with {\it AstroSat} has enabled the capture of crucial early outburst phases. These phases often exhibit unique signatures in disk geometry and coronal properties that are key to understanding state transitions. 

However, rapid rescheduling constraints often limit the ability to capture the very early stages of these outbursts where disk instability mechanisms are most active. To address these challenges, future observational strategies will require wide-field monitoring instruments in the soft-to-medium energy X-ray bands coupled with fast-response capabilities. Such systems would complement the strengths of {\it AstroSat} and XB-NEWS, expanding the observational window for studying transient behavior. These efforts will refine our understanding of transient evolution, the physics of disk instabilities, and the role of accretion dynamics in shaping the behavior of black hole binaries. 

Simultaneous X-ray and UV broadband observations of bright accretion disks will shed light on the reprocessing scenarios. High resolution UV spectroscopy in NUV, FUV bands can reveal the signatures of accretion disk-winds. Currently, {\tt UVIT} onboard {\it AstroSat} allows us to probe the signatures of accretion disk-wind. Moreover, a large area ($\sim 1$m$^{2}$) UV-spectrometer will complement the wide-band polarimetry and spectroscopy, and thus will be highly beneficial. 

A new promising field is the study of X-ray polarization. Broadband X-ray polarization measurements provide an additional dimension to our understanding of black hole systems by revealing their geometry and physical conditions. This extra layer of analysis enables us to probe the orientation and distribution of magnetic fields and scattering regions in these extreme environments. By comparing polarization contents across different energy bands it is possible to disentangle the relative contributions of the accretion disk, corona, and jet to the observed X-ray emission. India is leading this effort with results from {\tt CZTI}, and the field will get enhanced by {\tt POLIX} instrument on-board recent launch of the Indian satellite {\it XPoSat}. However, there is an urgent need to train young Indian astronomers in polarization analysis and perhaps, more importantly, in modelling the polarization signals.

\section{Future Growth}
\label{sec:futureplans}
Over the next couple of decades, with the growth of the Indian economy, it is expected that our country will have a larger scientific community and a number of mega-science projects. In these coming two decades, the world will witness several large-scale observational facilities, e.g., {\it LIGO-India Observatory}, {\it Einstein Telescope} ({\it ET}), {\it Cosmic Explorer} ({\it CE}), {\it DECIGO} ({\it DECIGO}), {\it Laser Interferometer Space Antenna} ({\it LISA}), and possibly a few other ground-based and satellite-based gravitational wave detectors accompanied by numerous multi-wavelength missions, e.g., {\it Square Kilometre Array} ({\it SKA}), {\it Large Synoptic Survey Telescope} ({\it LSST})/{\it Vera Rubin Observatory}, {\it Thirty Meter Telescope} ({\it TMT}), {\it Advanced Telescope for High-ENergy Astrophysics} ({\it ATHENA}), {\it Daksha}, etc., which will be transformative for the astronomical science. India also has significant involvement in several of these future observational facilities. It is rather natural that the Indian share of contributions in this global effort of astronomical science will also grow considerably.

During these next two decades of Indian astronomy, the community is expecting the advancement of large and modern astronomical facilities in different observational windows of electromagnetic and gravitational wave spectra, high-energy neutrinos, and cosmic rays, a truly remarkable combination of multi-messenger astronomy~\citep{gw_em_nu_whitepaper}. This will open up a whole new spectrum of opportunities and scope of scientific explorations to address several outstanding questions related to compact objects, viz. neutron stars and black holes. This will, in turn, make it possible to get crucial inputs from multi-messenger astronomy to our understanding of several key aspects in fundamental nuclear interactions at ultra high-density physics, the behaviour of strong nuclear force, and various unexplored physics of the matter under extreme conditions which have been a subject of mostly theoretical speculations to date. 

Understanding the structure and composition of neutron stars is a multi-disciplinary effort. It requires a joint venture by different communities including nuclear physics, computational fluid dynamics, multi-messenger observational astronomers, modern data analysts and statisticians. Computational physics has started playing a crucial role in different areas of modern physical sciences in recent times, including astronomy and astrophysics. With the advancement of computational effort globally in addressing fundamental scientific problems, many of the theoretical modelling and simulations which have so far been challenging, are now possible to achieve. 

In parallel, there has been a strong drive in large-scale information processing and data-science-driven statistical inference to complement the progress in theoretical modelling and predictions. In the next couple of decades, this combined effort on both theoretical and observational fronts augmented by computational techniques has a great promise to drive the next generation of astronomy and astrophysics to a successful endeavour. In this context, the Indian astronomy community should take a more inclusive approach and broaden its reach to make connections with multi-disciplinary sciences and technological advancements to fulfill its future goal. 

\subsection{Key Research Problems for the Future}
We can briefly summarize the state of the field and the open directions of interest to the community below.
\subsubsection{White Dwarfs}
    \begin{enumerate}
        \item Magnetic field interactions, internal \& external; Effects on cosmology through Type Ia supernovae, chemical abundances through novae
        \item Common envelope and binary evolution as a population; Effects on binary mergers, binary fractions, and GW detections
        \item Emission mechanisms of WD - companion interactions
    \end{enumerate}

\subsubsection{Neutron Stars}
\begin{enumerate}
    \item Equation of state of dense matter.
    \item Emission Physics; Nulling, glitching and dynamic behaviour.
    \item Evolutionary pathways and interconnections between families of neutron stars.
    \item Binary evolution and population synthesis; Effects on LIGO detections, GAIA astrometric signatures
    \item Behaviour and effects of magnetic fields; on neutron star mass, oblateness, ellipticity, continuous gravitational wave signals
\end{enumerate}
\subsubsection{Black Holes}
\begin{enumerate}
    \item Binary evolution and population synthesis; Effects on LIGO detections, GAIA astrometric signatures
    \item Effects of spin and accretion processes on jets and jet launching
    \item Physical understanding of QPOs from accretion disks
    \item Radiative behaviour of accretion disks
    \item Connection between the associations of different types of QPOs with the radio jets having variable strengths 
\end{enumerate}

\subsection{Existing Observational Facilities}

\subsubsection{Upgraded GMRT and Expanded GMRT: } 

The potential of {\it GMRT} in domain of searching for pulsars remained untapped for a long duration. Up-coming surveys of pulsars with a parameter space (with high-resolution observations, and compute-intensive data analysis) that was not explored before, promise the discovery of a bunch of pulsars, millisecond pulsars, long-period pulsars, and rotating radio transients. In addition, with the unprecedented sensitivity that will be available with {\it Expanded GMRT} ({\it eGMRT}), all the searches and follow-up investigations of pulsars will take a tremendous leap. 

\subsubsection{X-ray and Multi-wavelength Satellite AstroSat: } 

It is India’s first dedicated multi-wavelength astronomical observatory, launched by PSLV on 28 September 2015. {\it AstroSat} carries five scientific payloads: {\tt UVIT}, {\tt SXT}, {\tt LAXPC}, {\tt CZTI} and {\tt SSM}, and they are capable of simultaneous observations from ultraviolet to Gamma X-rays. It has completed 9 years of in-flight operations in September 2024 as a proposal-based observatory. Currently, it has close to 1500 global users (nearly 50\% of them are Indian users) and has resulted in more than 480 articles in peer-reviewed journals, more than 1700 conference proceedings, telegrams, circulars, and other non-refereed publications. 

From stellar to extragalactic astrophysics, a large number of scientific contributions are made. The detection of FUV emission from Her X-1, cyclotron line from few NSXBs, burst oscillations and QPOs, ranging from mHz to kHz, spin-estimation of black holes, and detailed study of type-I X-ray bursts, are few notable. Apart from the high volume of scientific output, {\it AstroSat} reached a milestone by exceeding the expected lifetime of 5 years proposed during the conception of the mission. Despite the fact that all instruments were built more than 15 years ago, most of them are still performing surprisingly well: except for issues like telemetry drops and gaps in X-centroids in CCD images, {\tt SXT} and {\tt CZTI} instruments, the FUV channel of {\tt UVIT} and one of the {\tt LAXPC} units are performing well and producing scientific results.

%Use of existing facilities/instrumentation upgrades.
%Would INSIST be good for compact objects?

%gamma-/X-ray/Astrosat --> Sudip 

%Optical/IR --> Gulab (AGN), Shriharsh (neutron stars)
% small telescope use -- ST
%\subsubsection{High Speed Optical/IR Spectrophotometry}
%1. Ultrafast optical and IR cameras with a frame rate of 500 Hz to 1 kHz and low-noise X-ray detectors with a time resolution of microseconds. They need to be working simultaneously. Ultracam (CA but not in India), Silicon drift detector arrays (YB)
%2. We need detector with the background of < 10 milliCrab, cooled SDD is ideal, working in 2-30 keV (YB),  Chandra (CA)

%GeV/TeV/PeV -->Pratik Majumdar (CTA, MACE) Indra.
%\subsubsection{Stereoscopic MACE}

\subsection{Theoretical Studies \& Computational Facilities}

% Indra
% Existing facilities + future facilities
% Future technologies (hardware acceleration, AI/ML).
% 

Multiple research directions involving compact objects need computational tools — including searches and characterizations of gravitational waves, radio, and X-ray pulsar signals, simulations of GRMHD, nuclear interactions at different density regimes, thermonuclear
reactions in hot accretion flows and supernovae, the evolution of compact object populations, etc. — require vast amounts of computational resources and skills to use them. A detailed view of computational astrophysics has recently been produced by the Indian computational astrophysics community~\citep{comp_astro_AsiVisionDoc2025}. Here we discuss several key aspects relevant to astrophysics with compact objects. 

Indian community is actively building new software and releasing those for the use of other researchers. One example is ‘PINTA’, a software developed by the Indian Pulsar Timing Array consortium to analyse uGMRT data~\citep{sushobhanan2021} that has been publicly released as an observatory facility~\footnote{The downloadable link can be found from the NCRA website (http://www.ncra.tifr.res.in/ncra/gmrt/gmrt-users/pinta).}. As a community, we need to develop large-scale distributed computing facilities that are accessible across the country and competitively allocated to researchers both in research institutes and in universities. We also need more training initiatives along the lines of hands-on computing workshops, online tutorials, and support staff to help students, postdocs, and faculty members to efficiently utilise these facilities.   

Gravitational wave astronomy has started in the past decade but is maturing with routine detections of new transient events that essentially involve compact objects -- black holes and neutron stars~\citep{lvk_gwtc3_2023PhRvX}. Discovery of these sources has paved new ways to explore several aspects of fundamental physics~\citep{lvc_tgr_gwtc1_prd2019, lvk_tgr_gwtc3_2021arXiv211206861T, gw170817_eos_2018, lvc_tgr_gw170817_2016ApJL} and astrophysics~\citep{lvk_pop_bbh_2023PhRvX, lvk_rate_bbh_2016ApJ, lvc_2019ApJ_876L_7S}. The LVK Collaboration is also searching for signals from the continuous gravitational wave sources, stochastic GW backgrounds of both cosmological origin and combinations of unresolved astrophysical sources. Most of these searches as well as their characterizations and source property studies require a significantly large amount of computational resources~\citep{brady_etal_prd1998, MPW_PRD2023}, in the form of CPU, GPU and skilled persons with technical knowledge. Building up national facilities for high-performance computing (HPC) supported by essential skilled professionals will be desirable to make this endeavor a success. 

In parallel, the community needs to develop more sophisticated numerical magneto-hydrodynamic codes (though codes like BHAC, H-AMR, and H-AMRAD, are available), which would require high-performance computing facilities to be made available to the community. Some of the publicly available simulation codes used to study accretion onto black holes, are BHAC, Athena++, HARMPI, etc. BHAC is built on the MPI-AMRVAC framework. It solves ideal, GRMHD equations of motions in 3+1 dimensions, the choice of solvers are TVDLF and HLL. By using the Athena++ one can work in non-relativistic, special relativistic as well as GR HD and MHD scenarios. A wider variety of solvers are provided although generally HLLC is used for HD and HLLD for MHD. However, only for fixed adiabatic index equation of state is available. HARMPI solves GRMHD equations in Kerr-Schild spacetime, and uses HLL and LLF solvers, but has no static or adaptive mesh refinement. Therefore, various codes specialize in highlighting different aspects of flow motions around black holes. There is no single version of code that can solve all possible problems.  

The majority of the publicly available numerical computation tools and software contain enormously large codebases and it is almost impossible for an independent researcher to implement newer physical processes in the code. Furthermore, developing simulation codes are active and evolving area of research. The last word has not been written about the best scheme or the best reconstruction method. As a result, newer and more efficient codes are being developed now and will continue in the future. At present, there is a large community of young researchers in India, who are being trained in code development. Most of the accretion models being tested abroad are those that regenerate the Keplerian disc or the ADAF, or recently the MAD and SANE configurations. Very little effort goes into the studies of advection and accretion. By participating in this active area of research through collaborations and individual efforts, one can benchmark the code parameters to test alternative accretion models along with the ones that are currently being tested elsewhere. Exploring the option of developing newer codes with specific routines we will get more freedom to develop and maintain it in a way that is better suited for the Indian community and can utilize available computational infrastructure, e.g., CPUs and GPUs.

In modern times, GPU-based simulations are becoming an increasingly popular choice due to their computationally efficient and stable operation. For example, if a CPU-based simulation takes 15.4 seconds per time-step on 32 cores, a GPU-accelerated simulation can potentially bring this down to a few seconds per time-step, depending on the hardware. The GPU requirements for a 3D GRMHD accretion disk simulation depend heavily on the resolution, domain size, and complexity of the simulation. For small-to-medium grids (e.g., 512$^3$ to 1024$^3$), modern GPUs like the RTX 4090 or A5000/A6000 with 24–48 GB of Video Random Access Memory (VRAM) should suffice. For larger simulations (e.g., 2048$^3$ or more) or long-duration, high-resolution simulations, we will need high-memory GPUs, e.g., NVIDIA A100 or H100, or possibly even multi-GPU setups to handle the demanding workloads. Some advanced MHD codes (such as PLUTO, Athena++, or RAMSES) already have GPU support built-in, utilising libraries like CUDA (for NVIDIA GPUs) or HIP (for AMD GPUs).  

A new rapidly changing aspect of computation is advanced accelerators. Apart from graphics processing units (GPUs), field programmable gate arrays (FPGAs), tensor processing units (TPUs), neuromorphic computing units\footnote{For example:~\href{https://www.intel.com/content/www/us/en/research/neuromorphic-computing.html}{see Intel's neuromorphic computing} } are now being integrated into computing nodes to form heterogeneous clusters. Computational astrophysicists, in collaboration with computing centres such as CDAC may invest time and effort into understanding the applications of these technologies and how they can boost simulation and computational astrophysics codes.

%Most students coming into graduate programs across the country are under-prepared for computational and experimental/instrumentation-oriented work. While programming is nominally taught in undergraduate physics courses, this usually entails writing small scripts and using plotting libraries, which is considered adequate, although some of the institutes have courses on computational techniques dedicated to Ph.D. students, even in their respective department of physics/astronomy. However, in computationally demanding research fields, the complexity and scale of code have grown significantly to rival code-bases in industry. For example, the Modules for Experimental Stellar Astrophysics (MESA; \url{https://github.com/MESAHub/mesa}), Cambridge stellar evolution code STARS (https://people.ast.cam.ac.uk/~stars), GRMHD accretion disk/jet code HARMPI (https://github.com/atchekho/harmpi) have a few million lines of \texttt{FORTRAN} and \textit{C}. In order for our students to use these code bases and successfully contribute to new ones, we need a systematic scientific programming curriculum that incorporates industry standard methodologies and tools --- version control, continuous testing, containerization, etc. Companies offering cloud computing services, etc., appear to be interested in funding code-developing activities. This area can be explored in view of our requirements and expertise in the area of heavy computing. As an additional option, this initiative will also assist to train the students for jobs in industrial research and other sectors outside of core academia. 

\vspace{1cm}

\subsection{Community Development}
The bedrock of any research community is its students and young researchers. Thus, it is imperative that we attract and train young minds. While compact objects (especially pulsars and black holes) are extremely well-popularized in popular media, most students are not exposed to real research in these areas and are not trained in the tools and techniques required for it. 

%It will be beneficial if various research groups and institutions across the country come together and build synergistic research programs that work towards a common goal. Presently, we have a very successful research programme, Joint Astronomy Programme (JAP), originally initiated to develop a research network via students’ involvement including many institutes across the country, but the participating institutes have recently been limited to only those based in Bangalore due to various logistics. One of the plans could be to revamp JAP to include many institutes spread across the country and create a broader platform for the Astronomy and Astrophysics graduate school. However, an important aspect of it is the requirements of financial resources. 

%Apart from the funding towards teaching infrastructure, it should provide the essential support for other academic activities e.g., schools, workshops, conferences and academic visits of eminent scientists and researchers from across the country and from abroad. Funding agencies and research institutions should encourage competitively-resourced long-term collaborations through specialized streams that encourage (a) multi-year collaborations with well-identified and well-justified goals, (b) the development, growth, and sustenance of these collaborations through funding for visits, workshops, and resource development (e.g., dedicated tutorials/documentation). 

We consider the following points for strengthening and accelerating sustained community development: 

\begin{enumerate}
    \item \textbf{Curricula and shared tutorials/online courses for introducing undergraduate students to research:} The Indian astronomy community associated with different aspects of compact object science can develop shared and maintained tutorials that undergraduate or early graduate students can undertake independently to learn the skills required for different research projects. For some of these curricula, we propose to bring in expertise from and collaborate with other communities such as nuclear, condensed matter, statistical physics, etc., on the theoretical front.  
    
    On the observational side, specialised training for radio, X-ray and gravitational wave astronomical observations will be necessary. Specifically, in the era of Indian effort in the SKA project, some key steps have been taken. An online repository for SKA users curated by the engineering team and science working group members is planned to benefit increasing exposure related to SKA and SKA-related compact object science among students~\citep{SKAepo}. Similarly, in the context of GW observations, IUCAA, along with other institutions such as IIT Bombay and ICTS, hosts in-person study hubs following the Gravitational Wave Open Data workshop (GWODW) of the LIGO-Virgo-KAGRA (LVK) collaboration every year. The workshop and study hubs tackle topics such as access to LVK public data, introduction to gravitational wave (GW) detectors and their data, techniques for GW data analysis, and interpretation of compact binary merger signals. 
    
    \item \textbf{Annual astronomy schools spanning 3-4 weeks to a couple of months:} The Radio Astronomy Winter School (RAWS), held annually in winter and spanning two weeks, is organised jointly by IUCAA’s Astronomy Centre for Educators (ACE) and NCRA for faculty and students of colleges and universities. The participants get introduced to radio astronomy through a series of lectures, demonstrations, and hands-on experiments. In the past, several {\it ASTROSAT} workshops on pulsar and black hole astrophysics, along with X-ray timing and spectroscopy, were conducted by TIFR, BARC, and IUCAA. Visiting Student Program (VSP) of Indian Academy of Science (ASI) also supports several students for projects in Astronomy and Astrophysics in appropriate departments/institutions, and RRI also has several student projects. The primary aim of this workshop series was to familiarize M.Sc. and Ph.D. students, post-doctoral fellows, and other interested astronomers in India with the {\it ASTROSAT} mission’s objectives and capabilities and to expand the astrophysical community which can carry out scientific research using {\it ASTROSAT}. 

    \item \textbf{Instrumentation \& experimental methods:} A suitable group of students can be trained with the instrumental and experimental perspectives, for example, (a) they may start with instrumentation projects like working with telescopes and cameras, working with detectors and relevant electronics, understanding detector background, and ways to reduce them, (b) they can be trained for time series analysis with already available data in the archive from fast detectors around the world. Also, new techniques are needed to be developed to improve data and result qualities. In close association to this, different data challenge events can be planned for building up relevant expertise, especially keeping in mind the prospects with SKA, LIGO-India and other relevant observational facilities with significant involvement by the community.

    \item \textbf{Theory-observation-instrumentation workshops \& meetings:} There is an urgent need for more detailed interactions between theorists, observers, and instrumentation groups so that the gap between theory and observables can be bridged and future missions can be designed for the greatest impact. Recently, some effort has been initiated to foster an Indian community centered around multi-messenger astronomy to provide a platform for cross-disciplinary interactions and activities~\citep[see ][]{gw_em_nu_whitepaper}. 

    \item \textbf{Provide a flexible funding mechanism for short-term research students:} 
    Undergraduate students from science and engineering backgrounds, who intend to pursue either a university-credited project at M.Sc./post-M.Sc. internships, or at an equivalent level may be encouraged to take up short-term research projects at a suitable research lab/institute. This should be in addition to the existing regular summer student fellowships and other programs, and with the financial support to cover their essential needs. The VSP by ASI is an endeavour in this direction which could plausibly be scaled up to support larger members across the country. 

    \item \textbf{Citizen science projects \& competitive events:} The LIGO-Virgo-KAGRA (LVK) network provides many opportunities for science enthusiasts to take part in research in gravitational wave science through its diverse Citizen Science projects such as Gravity Spy, Black Hole Hunter, Black Hole Master, Black Hole Pong, and Pocket Black Hole. Similar tools can be developed (similar to citizen science projects) in other areas of astronomy so that school or college communities can use them easily and participate in activities relevant to research. For example, designing prototypes of real research problems (from theory/data/instrumental area) that can be used to identify possible paths by organising competitive events. The Astronomers in Residence program is an initiative by the ASI's Public Outreach and Education Committee which aims at these activities in a small-scale. 

    \item \textbf{Synergies with other working groups:} Compact objects are central to transient astronomy related to multi-wavelength electromagnetic channels, gravitational waves, and neutrino observations. Naturally, there are synergies with several other areas of astronomy including EM transients~\citep{vision_doc_em_transients}, GW transients and testing theories of gravity~\citep{vision_doc_gw_tgr}, supermassive BHs, etc. Many of these areas are covered by different working groups of the Astronomical Society of India (ASI). 

    \item \textbf{Connections outside of astrophysics:} Connections outside of astrophysics: The goal is to connect laboratory experiments (nuclear, hypernuclear, heavy-ion collisions, etc.) to astrophysics through consistent models of neutron stars and imposing constraints using multi-messenger (multi-wavelength electromagnetic, gravitational wave) observational data. Collaboration with other academic communities (e.g., high-energy physics, condensed matter physics, etc., as mentioned above) has the potential to bring new ideas and methods into the field. 
    
\end{enumerate}

\section*{Acknowledgement}

We are thankful to an anonymous referee for her/his advice which has significantly improved the overall quality of the manuscript. 
AB acknowledges SERB (SB/SRS/2022-23/124/PS) for financial support. AM acknowledges support from the DST-SERB Start-up Research Grant SRG/2020/001290. 

This document is coordinated and compiled by Arunava Mukherjee, Indranil Chattopadhyay and Shriharsh Tendulkar.

\bibliography{compact_objects}{}

\end{document}